\documentclass[onecolumn,showpacs,preprintnumbers,amsmath,amssymb]{revtex4-1}

\usepackage{graphicx}% Include figure files
\usepackage{dcolumn}% Align table columns on decimal point
\usepackage{bm}% bold math
\usepackage{color}

\def\vc#1{\mbox{\boldmath $#1$}}

\begin{document}

\preprint{}

\title{Weinberg operator contribution to the nucleon electric dipole moment in the quark model}% Force line breaks with \\

\author{Nodoka Yamanaka$^{1,2,3}$}
\email{nodoka.yamanaka@riken.jp}
\author{Emiko Hiyama$^{3,4,5}$}
\email{hiyama@phys.kyushu-u.ac.jp}
\affiliation{$^1$Amherst Center for Fundamental Interactions, Department of Physics, University of Massachusetts Amherst, MA 01003, USA}
\affiliation{$^2$Department of Physics, Kennesaw State University, Kennesaw, GA 30144, USA}
\affiliation{$^3$Nishina Center for Accelerator-Based Science, RIKEN, Wako 351-0198, Japan}
\affiliation{$^4$Department of Physics, Kyushu University, Fukuoka, 819-0395, Japan}
\affiliation{$^5$Advanced Science Research Center, Japan Atomic Energy Agency, Tokai, Ibaraki, 319-1195 Japan}

\date{\today}% It is always \today, today,

\begin{abstract}
We evaluate the contribution of the CP violating gluon chromo-electric dipole moment (the so-called Weinberg operator, denoted as $w$) to the electric dipole moment (EDM) of nucleons in the nonrelativistic quark model.
The CP-odd interquark potential is modeled by the perturbative one-loop level gluon exchange generated by the Weinberg operator with massive quarks and gluons.
The nucleon EDM is obtained by solving the nonrelativistic Schr\"{o}dinger equation of the three-quark system using the Gaussian expansion method.
It is found that the resulting nucleon EDM, which may reasonably be considered as the irreducible contribution, is smaller than the one obtained after ``$\gamma_5$-rotating'' the anomalous magnetic moment using the CP-odd mass calculated with QCD sum rules.
We estimate the total contribution to be $d_n = w \times 20 \, e \, {\rm MeV}$ and $d_p = - w \times 18 \, e \, {\rm MeV}$ with 60\% of theoretical uncertainty.
\end{abstract}

%CP invariance, 11.30.Er
%Nonrelativistic quark model, 12.39.Jh
%Models beyond the standard model, 12.60.-i
%Electric and magnetic moments, 13.40.Em

\maketitle

\section{Introduction}
\label{sec:intro}

The discovery of the Higgs boson in LHC experiments \cite{Aad:2012tfa,Chatrchyan:2012ufa} completed the particle listing of the standard model (SM), but it is known that still several cosmologically important phenomena cannot be explained.
One of them is the matter abundance of the Universe, which, under the criteria of Sakharov \cite{Sakharov:1967dj}, revealed us that the CP violation is crucially lacking \cite{Farrar:1993hn,Huet:1994jb}.
This is why the search of new sources of CP violation beyond the SM is currently so actively pursued.

The electric dipole moment (EDM) \cite{He:1990qa,Bernreuther:1990jx,khriplovichbook,Pospelov:2005pr,Fukuyama:2012np,Engel:2013lsa,Yamanaka:2014mda,Hewett:2012ns} is a very sensitive experimental probe of CP violation, and it may be measured in many systems.
Recently, the experimental result of the EDM of neutron was updated ($|d_n | < 1.8 \times 10^{-26}e$ cm) \cite{Abel:2020gbr}, and the constraint on the hadronic CP violation became tighter.
The measurements of the EDM of diamagnetic atoms \cite{Graner:2016ses,Bishof:2016uqx,Allmendinger:2019jrk} also constrain the neutron EDM at a similar level \cite{Ginges:2003qt,Yamanaka:2017mef,Chupp:2017rkp,Yanase:2020agg}.
Regarding the EDM of the proton, experimental measurements using storage rings are currently in development, with impressive prospective sensitivity of $O(10^{-29})e$ cm \cite{Farley:2003wt,Anastassopoulos:2015ura,bnl}.
There are also attempts to measure the EDM of flavored baryons in high energy accelerator experiments \cite{Aiola:2020yam}.
It is known that the nucleon EDM receives contribution from many CP violating quark-gluon level processes.
The most well-known one is the quark EDM, for which the systematics is now quite well controlled thanks to extensive lattice analyses \cite{Yamanaka:2018uud,Gupta:2018lvp,Alexandrou:2019brg,Cirigliano:2019jig,Horkel:2020hpi}, although there are still discussions regarding the disagreement with perturbative QCD extractions \cite{Bacchetta:2012ty,Kang:2015msa,Benel:2019mcq,DAlesio:2020vtw,Anselmino:2020nrk}.
Contributions from purely QCD operators such as the chromo-EDM or the $\theta$-term are also well studied in chiral effective field theory \cite{Crewther:1979pi,Pich:1991fq,Borasoy:2000pq,Mereghetti:2010tp,deVries:2010ah,Mereghetti:2010kp,deVries:2012ab,Fuyuto:2012yf,Mereghetti:2015rra,deVries:2015gea}, and there are also available lattice QCD results for the latter one \cite{Abramczyk:2017oxr,Dragos:2019oxn}, while the ab initio calculation of the former one is more challenging \cite{Rizik:2020naq}.

\begin{figure}[htb]
\begin{center}
\includegraphics[width=4cm]{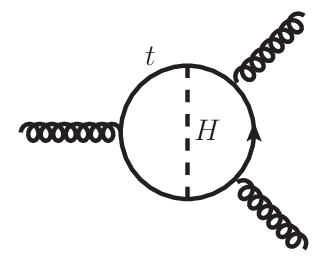}
\caption{\label{fig:weinberg_higgs}
Typical Feynman diagram contributing to the Weinberg operator in Higgs-doublet models.
}
\end{center}
\end{figure}

As a more obscure CP-odd strong interaction, we have the Weinberg operator \cite{Weinberg:1989dx,Dine:1990pf,Braaten:1990zt,Chemtob:1991vv}.
This is a dimension-6 CP-odd interaction defined by
\begin{eqnarray}
{\cal L}_w 
&=& 
\frac{1}{3!} w 
f^{abc} \epsilon^{\alpha \beta \gamma \delta} G^a_{\mu \alpha } G_{\beta \gamma}^b G_{\delta}^{\ \ \mu,c}
,
\label{eq:weinberg_operator}
\end{eqnarray}
where $f^{abc}$ is the structure constant of the $SU$(3) Lie algebra. 
It was first introduced by Weinberg as a potentially large contribution of the CP violation of the Higgs sector to the neutron EDM \cite{Weinberg:1989dx}.
The Weinberg operator is relevant in many candidates of new physics beyond standard model such as the Higgs-doublet models \cite{Weinberg:1989dx,Dicus:1989va,Boyd:1990bx,Cheng:1990gg,Bigi:1991rh,Bigi:1990kz,Hayashi:1994xf,Hayashi:1994ha,Wu:1994vx,Jung:2013hka,Brod:2013cka,Dekens:2014jka,Cirigliano:2016nyn,Cirigliano:2016njn,Panico:2017vlk,Cirigliano:2019vfc,Haisch:2019xyi,Cheung:2020ugr}, supersymmetric models \cite{Dai:1990xh,Arnowitt:1990je,Abel:2001vy,Demir:2002gg,Demir:2003js,Degrassi:2005zd,Abel:2005er,Ellis:2008zy,Li:2010ax,Zhao:2013gqa,Sala:2013osa,Hisano:2015rna}, and other models \cite{Chang:1990sfa,Rothstein:1990vd,Xu:2009nt,Choi:2016hro,Abe:2017sam,Dekens:2018bci,DiLuzio:2020oah}.
The SM contribution generated by the CP phase of the CKM matrix \cite{Kobayashi:1973fv} is negligibly small due to suppression by the GIM mechanism \cite{Booth:1992tv,Pospelov:1994uf,Yamaguchi:2020dsy}.
An example of the process generating the Weinberg operator in the Higgs-doublet model is shown in Fig. \ref{fig:weinberg_higgs}.

\begin{figure}[htb]
\begin{center}
\includegraphics[width=12cm]{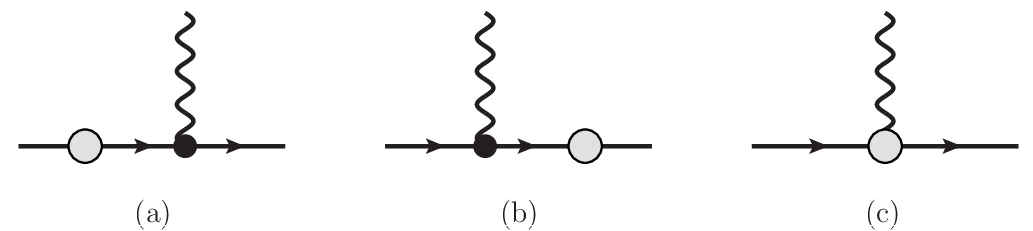}
\caption{\label{fig:nucleon_EDM_Weinberg_operator}
Diagrammatic representation of the nucleon EDM in the hadron level effective field theory.
Diagrams (a) and (b) are the contribution from the ``$\gamma_5$-rotated'' anomalous magnetic moment (black blob) by the CP-odd baryon mass (grey blob).
Diagram (c) is the irreducible nucleon EDM contribution, evaluated in this work.
}
\end{center}
\end{figure}

The analysis of the nucleon EDM generated by the Weinberg operator involves sizable theoretical uncertainty.
The most important obstruction in the quantification of its contribution to the nucleon level CP violation is the absence of quark field in Eq. (\ref{eq:weinberg_operator}) which implies that chiral techniques are not very powerful. Nevertheless, there were several attempts to quantify this effect.
The contribution of the Weinberg operator to the neutron EDM was treated in the original work of Weinberg \cite{Weinberg:1989dx} using the {\it na\"{i}ve dimensional analysis}, giving
\begin{equation}
d_N (w) 
\approx
e\frac{\Lambda}{4\pi}w
\approx
w \times 90 \, e \, {\rm MeV}
,
\label{eq:NDA}
\end{equation}
where $\Lambda = 4\pi f_\pi$.
There is also another work which used the same method to estimate the quark EDM \cite{Dib:2006hk}.
The na\"{i}ve dimensional analysis is of course not more accurate than the order estimation, and more precise ways to obtain the relation between $w$ and the nucleon EDM are definitely required.
Bigi and Uraltsev then inspected it in more detail at the hadron level, and came out that the neutron EDM is generated by the one-particle reducible and irreducible parts (see Fig. \ref{fig:nucleon_EDM_Weinberg_operator}) \cite{Bigi:1991rh,Bigi:1990kz}.
The reducible contribution is obtained by ``$\gamma_5$-rotating'' the anomalous magnetic moment with the CP-odd baryon mass.
The CP-odd mass was then evaluated using QCD sum rules, and the resulting nucleon EDMs were given as \cite{Demir:2002gg,Haisch:2019bml}
\begin{eqnarray}
d_N^{\rm (red)} (w) 
&\approx&
\left\{
\begin{array}{rl}
 w \times 25 \, e \, {\rm MeV} & (N = n ) \cr
-w \times 23 \, e \, {\rm MeV} & (N = p ) \cr
\end{array}
\right.
.
\label{eq:weinbergop_gpinn}
\end{eqnarray}
with 50\% of theoretical uncertainty \cite{Haisch:2019bml}.
The ideal method to obtain the coefficient relating $w$ and $d_N$ is lattice QCD, for which quantitative results are not yet available \cite{Cirigliano:2020msr,Rizik:2020naq}.
Another potentially interesting approach is to analyze higher twist contribution to the parton distribution functions \cite{Hatta:2020ltd}.

The analysis of QCD sum rules (\ref{eq:weinbergop_gpinn}) is however incomplete.
Indeed, the irreducible part has never been evaluated and has been neglected in previous studies, but there is no reason to omit it. 
The reducible ``$\gamma_5$-rotated'' contribution may in some sense be considered as the relativistic effect, since the CP-odd baryon mass and the anomalous magnetic moment are higher order terms in the velocity expansion.
On the other hand, the irreducible one has no reason to be generated from relativistic physics, and we may thus expect it to be derived from the evaluation of the internal structure in a nonrelativistic framework.
In this work we propose to calculate the Weinberg operator contribution to the nucleon EDM in the quark model.

The quark model was first conceived in the 60's by Gell-Mann to classify hadrons \cite{GellMann:1964nj}.
It was successful in describing hadron mass splittings, as well as other fundamental quantities such as the magnetic moment or the weak charge, using Lie algebra arguments by assuming that hadrons are $SU(3)_c$ singlet states composed of nonrelativistic (anti)quarks in the {\bf 3} ({\bf 3}$^*$) representation \cite{Sakita:1964qq,Beg:1964nm,Adler:1975he,Isgur:1979be,Capstick:2000qj}.
It is now considered to be one of the most successful model to describe low energy QCD, and the discovery of $\Omega$ in the early era \cite{Barnes:1964pd}, as well as that of double charmed baryons quite recently \cite{Aaij:2017ueg}, is demonstrating this.
The phenomenological interaction used is confining, i.e. the interquark potential becomes infinitely high at long distance and resembles the one-gluon exchange at short range.
Although the first principle derivation of the quark model from QCD is not available, lattice QCD calculations are strongly suggesting this picture \cite{Bali:1992ab,Takahashi:2000te}.
In view of the fact that the quark model describes well the excitations of hadrons \cite{Isgur:1978xj,Isgur:1978wd,Godfrey:1985xj,Capstick:1986bm}, the above phenomenological features are now undoubtedly established, and other aspects such as the interbaryon force at short distance \cite{Oka:1980ax,Oka:1981ri,Oka:1981rj} may also qualitatively well be derived.
The quark model is also working well for the description of hadron involving heavy quarks \cite{Roberts:2007ni,Karliner:2008sv,Yoshida:2015tia}, and it is also applied to the analysis of the pentaquarks \cite{Hiyama:2018ukv}, recently discovered by LHCb Collaboration \cite{Aaij:2015tga}.
In this way, the quark model is nowadays the first effective model of strong interaction to be used for testing or predicting hadronic quantities.
We note that some low lying baryons, such as the P-wave ones, are not reproduced by the quark model \cite{Isgur:1978xj}, which is probably due to the mixing with meson-baryon composite states.
We however anticipate that these states are not important in the study of the Weinberg operator contribution to the nucleon structure, since the creation of a pion by a purely gluonic operator suffers from chiral suppression.

The advantage of the quark model is that the wave function of the hadron may be calculated, and that physical quantities may systematically be evaluated within quantum mechanics.
The nucleons are considered as a nonrelativistic three-quark system with the phenomenological confining potential.
The calculation of the wave function of such systems is possible using the Gaussian expansion method \cite{Hiyama:2003cu} which is a powerful tool to solve few-body Schr\"{o}dinger equations.
This method has been applied in the calculations of the EDM of light nuclei \cite{Yamanaka:2015qfa,Yamanaka:2015ncb,Yamanaka:2016itb,Yamanaka:2016fjj,Yamanaka:2016umw,Lee:2018flm,Yamanaka:2019vec}, and we expect it to also work well for the present case of the nucleon EDM.
We will therefore evaluate the nucleon EDM generated by the Weinberg operator in the quark model with the expectation to obtain results that go beyond the accuracy of the na\"{i}ve dimensional analysis (\ref{eq:NDA}).
This will permit us to give an accurate constraint on the Weinberg operator from neutron EDM experiments.

We organize this paper as follows.
In the next section, we derive the one-loop level CP-odd gluon exchange potential between quarks generated by the Weinberg operator.
In Section \ref{sec:edm}, we present the setup of the calculation of the EDM of nucleons.
In Section \ref{sec:analysis}, we analyze our result by comparing it with the former result obtained using the QCD sum rules.
The final section is devoted to the conclusion.

\section{Quark level processes generated by the Weinberg operator\label{sec:quarklevel}}

In this section we calculate the CP-odd interquark force induced by the Weinberg operator.
It is generated by the one-loop level diagram of Fig. \ref{fig:2qint} (the Feynman rule for the three-gluon vertex is given in Appendix \ref{appendix:Feynman_rules}).
This diagram is diverging if we calculate it directly, so it is required to evaluate it with a scheme with a cutoff.
Here we use the heavy quark effective theory, which removes the relativistic degrees of freedom, corresponding to integrating out field configurations with energy-momentum component higher than the constituent quark mass $m_Q$.
This scheme is adequate for deriving interquark potentials that will be used in quark models.
The final form of the two-body scattering amplitude is
\begin{eqnarray}
i{\cal M}
&\approx &
w
\frac{N_c g_s \alpha_s m_g }{2 }
\frac{1}{|\vec{q}|^2 + m_g^2} 
\bar H_2 t_a H_2
\cdot
\bar H_1
t_a \vec{\sigma}\cdot \vec{q}
H_1
+(1 \leftrightarrow 2)
.
\label{eq:amplitude3}
\end{eqnarray}
where $q$ is the exchanged momentum, $H_i$ is the ``large'' component of the spinor of the $i$th quark, and $\alpha_s = 0.3$ is the QCD coupling chosen at the energy scale close to the nucleon mass.
We denote by $t_a$ the generator of the $SU(3)_c$ group.
The derivation is given in Appendix \ref{appendix:2-body}.
Here the mass of the gluon is given by $m_g \approx 350$ MeV, which we took from the gluon propagator calculation in Landau gauge lattice QCD \cite{Falcao:2020vyr}.
The Compton wave length of this gluon is $1/m_g \approx 0.6 $ fm which is quite compatible with the picture where gluons are confined in a proton with the radius $r_N \sim 0.8$ fm.
The velocity counting of this CP-odd two-body force is $O(v)$, due to $\sigma^{0 \gamma} q_\gamma \gamma_5$.
After Fourier transforming Eq. (\ref{eq:amplitude3}), we obtain the following nonrelativistic CP-odd 2-body force in the coordinate space
\begin{equation}
{\cal H}_{CPV}
=
-
\frac{N_c g_s \alpha_s m_g }{2 }
w
(\vec{\sigma}_1- \vec{\sigma}_2) 
\cdot
\vec{\nabla} \frac{e^{- m_g |\vec{r}_1 -\vec{r}_2| }}{4 \pi |\vec{r}_1 -\vec{r}_2|}
(t_a)_1 \otimes (t_a)_2
.
\label{eq:cpvhamiltonian}
\end{equation}
This interaction has exactly the same Lorentz structure as the phenomenological CP-odd nuclear force with $\omega$ meson-exchange \cite{Towner:1994qe,Liu:2004tq,Yamanaka:2015qfa,Yamanaka:2016umw,deVries:2020iea}.
The EDM of a nucleon, composed of three quarks, can therefore be calculated like the EDMs of $^3$He and $^3$H with the CP-odd one-pion exchange force.

\begin{figure}[htb]
\begin{center}
\includegraphics[width=4cm]{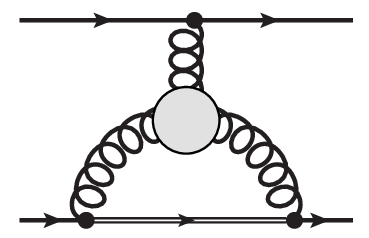}
\caption{\label{fig:2qint}
The one-loop level quark-quark amplitude generated by the Weinberg operator.
The thick lines denote the constituent quarks.
The double line is the dressed quark propagator which will be considered to be heavy.
}
\end{center}
\end{figure}

\begin{figure}[htb]
\begin{center}
\includegraphics[width=5cm]{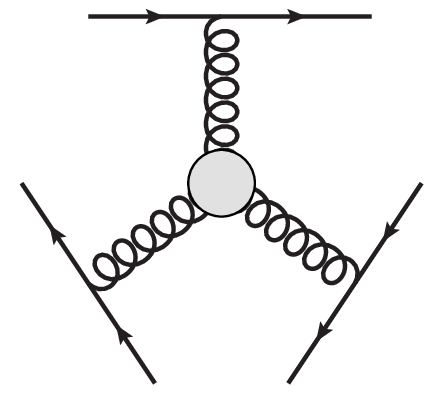}
\caption{\label{fig:3qint}
The three-quark amplitude generated by the Weinberg operator.
The thick lines denote the constituent quarks.
}
\end{center}
\end{figure}

The Weinberg operator also generates the CP-odd three-quark potential, through the three-quark amplitude depicted in Fig. \ref{fig:3qint} \cite{Blundell:2012ei}.
However, the momentum dependence of the Weinberg operator (see Fig. \ref{fig:wein_feyn} of Appendix \ref{appendix:Feynman_rules}) brings a velocity suppression of $O(v^3)$, so the CP-odd three-quark interaction should therefore be subleading.
In the calculation of the nucleon EDM, we therefore only consider the 2-body force (\ref{eq:cpvhamiltonian}) as the source of CP violation. 

\section{Calculation of the nucleon EDM\label{sec:edm}}

\subsection{Definition of the nucleon EDM}

Now that we have the CP-odd interquark force, we can evaluate the nucleon EDM.
The nucleon EDM is defined as
\begin{eqnarray}
d_{N}
&=&
\sum_{i=1}^{A} \frac{e}{2} 
\langle \, \Psi_N \, |\, (1/3 +\tau_i^z ) \, {\cal R}_{iz} \, | \, \Psi_N \, \rangle
\nonumber\\
&=&
\sum_{i=1}^{A} \frac{e}{2} 
\langle \, \Psi_N \, |\, \tau_i^z \, {\cal R}_{iz} \, | \, \Psi_N \, \rangle
,
\label{eq:nucleonedmpolarization}
\end{eqnarray}
where $|\, \Psi_N \, \rangle$ is the state of nucleon polarized along the $z$-axis, and $\tau^z_i$ is the isospin Pauli matrix.
The $z$-component of the coordinate of the $i$-th constituent quark ${\cal R}_{iz}$ is defined in the center of mass frame of the nucleon (that is why the isosinglet term with $1/3$ of the first equality does not contribute).
The nucleon state must be a mixing of opposite parity states to yield a nonzero nucleon EDM, which is possible if the CP-odd interquark force of Eq. (\ref{eq:cpvhamiltonian}) is present in the Hamiltonian.

\subsection{Quark model\label{sec:quarkmodel}}

In this work, we use the quark model to describe the CP-even strong interaction.
The nucleon is considered as a 3-body system of nonrelativistic constituent quarks.
We use the parameters of Refs. \cite{Bhaduri:1981pn,Semay:1994ht} which considered the interquark 2-body interaction.
The 3-body wave function is calculated with the Gaussian expansion method \cite{Hiyama:2003cu} in the coordinate space.

In this work we use three types of interactions.
The first one is the potential proposed by Bhaduri \cite{Bhaduri:1981pn}, given as
\begin{equation}
V_{qq , ab} (r)
=
-\frac{3}{16}
\lambda_a \lambda_b
\Biggl[
-\frac{\kappa}{r} 
+\lambda r 
+\Lambda
+\frac{\kappa}{m_Q^2} \frac{e^{- r /r_0}}{r r_0^2} \vc{\sigma}_a \vc{\sigma}_b
\Biggr]
,
\label{eq:bhaduri}
\end{equation}
where $m_Q = 0.337$ GeV, $\kappa = 0.52$, $\Lambda = - 0.9135$ GeV, $\lambda = 0.186 $ GeV$^2$, and $r_0 = 2.305$ GeV$^{-1}$ \cite{Bhaduri:1981pn}.
The other two proposed in Ref. \cite{Semay:1994ht}, named AL1 and AP1, are given by 
\begin{equation}
V_{qq , ab} (r)
=
-\frac{3}{16}
\lambda_a \lambda_b
\Biggl[
-\frac{\kappa (1-e^{-r/r_c})}{r} 
+\lambda r^p 
+\Lambda
+\frac{2 \pi \kappa'}{3 m_Q^2} (1-e^{-r/r_c}) \frac{e^{- r^2 /r_0^2}}{\pi^{3/2} r_0^3} \vc{\sigma}_a \vc{\sigma}_b
\Biggr]
,
\label{eq:semay}
\end{equation}
where $r_0 \equiv A / m_Q^B $.
The parameters are given in Table \ref{table:semay}.

\begin{table}[htb]
\caption{
Input parameters of the interquark potentials AL1 and AP1 \cite{Semay:1994ht}.
}
\begin{center}
\begin{tabular}{l|ccccccccc|}
Potential & $p$ & $r_c$ & $m_Q$ [GeV] & $\kappa$ & $\kappa'$ & $\lambda$ [GeV$^{1+p}$] & $\Lambda$ [GeV] & $B$ & $A$ [GeV$^{B-1}$] \\ 
\hline
AL1 & 1   & 0 & 0.315 & 0.5069 & 1.8609 & 0.1653 & -0.8321 & 0.2204 & 1.6553 \\
\hline
AP1 & 2/3 & 0 & 0.277 & 0.4242 & 1.8025 & 0.3898 & -1.1313 & 0.3263 & 1.5296 \\
\hline
\end{tabular}
\end{center}
\label{table:semay}
\end{table}

\subsection{Gaussian expansion method}

The object of the Gaussian expansion method is to solve the nonrelativistic Schr\"{o}dinger equation 
\begin{eqnarray}
( H - E ) \, \Psi_{JM,TT_z}  = 0 ,
\label{eq:schr7}
\end{eqnarray}
with the variational principle.
The Hamiltonian to diagonalize is
\begin{equation}
H
=
\sum_a T_a+ \sum_{a,b} V_{qq , ab}
+{\cal H}_{CPV , ab}
,
\label{eq:hamil7}
\end{equation}
with the kinetic energy operator $T$, the CP-even interquark potential $V_{qq , ab}$ seen in Sec. \ref{sec:quarkmodel}, and the CP-odd one ${\cal H}_{CPV , ab}$ (\ref{eq:cpvhamiltonian}).
Explicit indices indicating the labels of the constituent quarks $a$ and $b$ were written.

\begin{figure}[htb]
\begin{center}
\includegraphics[width=14cm]{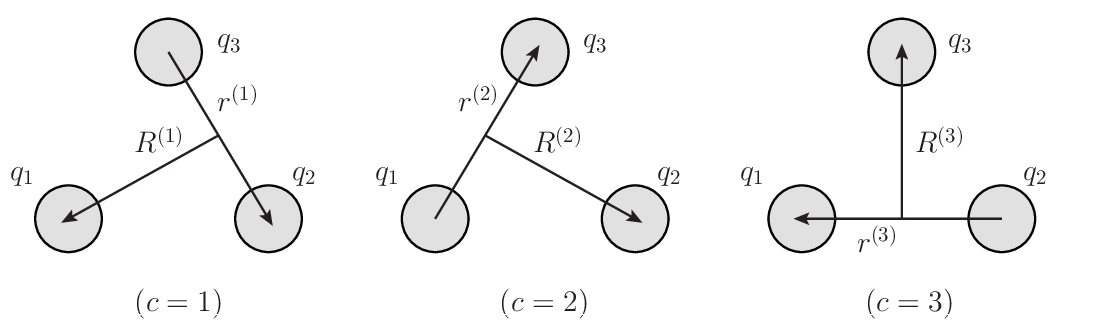}
\caption{\label{fig:jacobi}
The Jacobi coordinate for the 3-quark systems.
}
\end{center}
\end{figure}

We need two coordinates to express the wave function of three-quark systems.
The wave function of the 3-quark systems is given as the sum over three kinds of Jacobian coordinates shown in Fig. \ref{fig:jacobi}, as
\begin{eqnarray}
\Psi_{JM, I, I_z}(n, p)
&=&
 \sum_{c=1}^{3} \:
\sum_{nl, NL}
\sum_{T} \sum_{\Sigma} \sum_{s}
C^{(c)}_{nl,NL, \Sigma s, T}\: {\cal A} \Biggl[
\Bigl[  \eta^{(c)} ( T_c ) \otimes \eta'^{(c)} ({\scriptstyle \frac{1}{2}} ) \Bigr]_{I={\scriptstyle \frac{1}{2}} ,I_z}
\times 
\nonumber  \\
&&
\Bigl[
[ \phi^{(c)}_{nl}({\bf r}_c) \otimes \psi^{(c)}_{NL}({\bf R}_c)]_\Lambda \otimes \, \bigl[ \chi^{(c)} ( s_c ) \otimes \chi'^{(c)} ({\scriptstyle \frac{1}{2}} ) \bigr]_\Sigma \Bigr]_{J={\scriptstyle \frac{1}{2}}, M}
\times
| {\rm color\, singlet} \rangle
\Biggr]
,
\nonumber\\
\label{eq:he7lwf}
\end{eqnarray}
where ($\eta$,$\eta'$), ($\phi$,$\psi$), and ($\chi$,$\chi'$) are two isospin, radial, and spin components of the nucleon state according to the Jacobi coordinate, respectively.
Here $\cal{A}$ antisymmetrizes the whole system, i.e. the arguments inside the large bracket are entangled so as to be odd under the interchange of each quark.
Since the color part is completely antisymmetric, $l+s+T$ must be even.
The spin and isospin matrix elements can be calculated exactly in the same way as those of $^3$He and $^3$H nuclei, except that the antisymmetry requires $l+s+T$ to be even.
The radial components $\phi$ and $\psi$ are expanded with the following basis
\begin{eqnarray}
\phi_{nlm}({\bf r})
&=&
r^l \, e^{-(r/r_n)^2}
Y_{lm}({\widehat {\bf r}})  \;  ,
\nonumber \\
\psi_{NLM}({\bf R})
&=&
R^L \, e^{-(R/R_N)^2}
Y_{LM}({\widehat {\bf R}})  \;  ,
\end{eqnarray}
where the following geometric progression for the Gaussian range parameters is used:
\begin{eqnarray}
      r_n
      &=&
      r_{\rm min} a^{n-1} \quad 
      (n=1 - n_{\rm max}) \; ,
\nonumber\\
      R_N
      &=&
      R_{\rm min} A^{N-1} \quad
     (N \! =1 - N_{\rm max}).
\end{eqnarray}
In this work we use all angular momentum channels with $l, L, \Lambda \leq 2$, for which good convergence of the result is obtained.

To see the validity of the three-quark calculation, we display the up- and down-quark spin matrix elements of the nucleon obtained in the quark model.
The numerical values of the spin of the proton (neutron) are
\begin{equation}
\sigma_{p(n)} 
= 
(1.32-1.33) \sigma_{u(d)} - (0.32-0.33) \sigma_{d(u)}
,
\label{eq:Nedmquarkmodel}
\end{equation}
where the range is showing the variation due to the change of the interquark potential (Bhaduri, AL1, AP1, see Sec. \ref{sec:quarkmodel}).
This is in very good agreement with the analytic formula of the heavy quark limit $\sigma_{p(n)} = \frac{4}{3} \sigma_{u(d)} - \frac{1}{3} \sigma_{d(u)}$.
The small difference between them is due to the mixing of orbital angular momentum states.
We note that the contribution of the EDM of ``constituent quarks'' to the nucleon EDM is also given by the same coefficients, but the EDM of current quarks does not coincide with that of the constituent quarks due to additional effects such as the gluon dressing \cite{Yamanaka:2013zoa,Yamanaka:2014lva}.
Recent lattice studies \cite{Yamanaka:2018uud,Gupta:2018lvp,Alexandrou:2019brg,Cirigliano:2019jig,Horkel:2020hpi} are giving $d_{p(n)} \approx 0.8 d_{u(d)} - 0.2 d_{d(u)}$ which is smaller than Eq. (\ref{eq:Nedmquarkmodel}).

\section{Result\label{sec:analysis}}

From our quark model calculation, we obtain the following result:
\begin{eqnarray}
d_N^{\rm (irr)} (w) 
&\approx&
\left\{
\begin{array}{rl}
-w \times (4-5) \, e \, {\rm MeV} & (N = n ) \cr
 w \times (4-5) \, e \, {\rm MeV} & (N = p ) \cr
\end{array}
\right.
,
\label{eq:weinbergop_result}
\end{eqnarray}
where the range of the values indicates the uncertainty related to the choice of the interactions of the quark model (see Sec. \ref{sec:quarkmodel}).
The values are significantly smaller than the QCD sum rules outputs (\ref{eq:weinbergop_gpinn}) \cite{Demir:2002gg,Haisch:2019bml}.
As mentioned in the introduction, our result corresponds to the irreducible contribution of Fig. \ref{fig:nucleon_EDM_Weinberg_operator} (c), and it is consistent with the dominance of the EDM induced by the ``$\gamma_5$-rotated'' anomalous magnetic moment [Figs. \ref{fig:nucleon_EDM_Weinberg_operator} (a) and (b)] at the hadron level.
We note that the heavy quark approximation that we used in deriving Eq. (\ref{eq:amplitude3}) is not very accurate because the gluon mass is not too far from the constituent quark mass.
However, even with uncertainties of 100\%, the correction we calculated will not significantly affect the dominant contribution from the $\gamma_5$-rotated effect, in particular in light of the sizable uncertainty on the latter.
Keeping the uncertainty in mind, the total contribution to the nucleon EDM is 
\begin{eqnarray}
d_N (w) 
=
d_N^{\rm (red)} (w)
+ 
d_N^{\rm (irr)} (w) 
&\approx&
\left\{
\begin{array}{rl}
w \times 20 \, e \, {\rm MeV} & (N = n ) \cr
-w \times 18 \, e \, {\rm MeV} & (N = p ) \cr
\end{array}
\right.
.
\label{eq:weinbergop_total}
\end{eqnarray}
The theoretical uncertainty can be estimated by adding in quadrature the absolute values of our result (\ref{eq:weinbergop_result}) and the error bar 50\% for the $\gamma_5$-rotated contribution obtained in Ref. \cite{Haisch:2019bml}.
This amounts to about 60\% for both the neutron and the proton.

Let us derive the constraint on the Weinberg operator imposed by the recent experimental data of the neutron EDM.
The upper limit to $w$ at the scale $\mu = 1$ TeV is 
\begin{eqnarray}
|w (\mu =1 \, {\rm TeV})| < 2.3 \times 10^{-10} {\rm GeV}^{-2}
,
\end{eqnarray}
where the renormalization was calculated in the leading logarithm approximation \cite{Braaten:1990gq,Chang:1990jv,Degrassi:2005zd,Kamenik:2011dk,Dekens:2013zca,Sala:2013osa,Yamanaka:2017mef,deVries:2019nsu}.
Na\"{i}vely, we might see that the above equation is constraining the scale of new physics at the level of $O(100)$ TeV, but we have to note that the Weinberg operator only appears from the two-loop level in many known candidates of new physics \cite{Weinberg:1989dx,Abe:2017sam}, so that the constraint on the energy scale can be attenuated.

Let us try to interpret our result.
The reducible part calculated in Refs. \cite{Demir:2002gg,Haisch:2019bml} is obtained after the $\gamma_5$-rotation, so it is inheriting the large size of the anomalous magnetic moment \cite{Seng:2014pba}, which is an $O(1)$ quantity for nucleons.
Our calculation, in contrast, is encompassing the electromagnetic external field and the CP-odd interaction in a same quark level package, 
so it cannot be enhanced beyond the extent of the baryon.
Another argument is that the irreducible part must involve opposite parity excited states in the intermediate states, which require $O(100)$ MeV of energy transition.
Our result might also have overlap with the $\gamma_5$-rotated anomalous magnetic moment, but we estimate this effect to be small, since we calculated in a nonrelativistic framework, while the $\gamma_5$-rotation
is a relativistic one.

\section{Summary}

We calculated the CP-odd two-body force generated by the Weinberg operator and the nucleon EDM induced from it in the nonrelativistic quark model.
In the language of the velocity expansion, the CP-odd two-body force gives the leading contribution to the nucleon EDM.
We solved the 3-body problem in the coordinate space formalism using the Gaussian expansion method.
It was possible to calculate the nucleon EDM with the same calculational procedure as that of the EDM of $^3$He/$^3$H.
We obtained a smaller nucleon EDM than the one calculated with the QCD sum rules \cite{Demir:2002gg,Haisch:2019bml}.
We interpret this result as the dominance of the $\gamma_5$-rotated anomalous magnetic moment over the 
irreducible nucleon EDM.

Our analysis and results make us expect that the leading contribution to the isoscalar contact CP-odd nuclear force is also due to the $\gamma_5$-rotated CP-even contact $NN$ interaction determined in chiral effective field theory by the CP-odd nucleon mass.
This contact CP-odd nuclear force is one of the important subprocesses generating the nuclear and atomic EDMs.
For that we have to show again that the direct contribution to the isoscalar contact CP-odd nuclear force is small.
This analysis is also possible using the quark model, but it will be left for a future work.

\begin{acknowledgments}
We thank Jordy de Vries for useful discussions.
This work was supported by the Grant-in-Aid for Scientific Research (Grant No. 18H05407) from the Japan Society For the Promotion of Science.
\end{acknowledgments}

\appendix

\section{Feynman rules of CP-odd three-gluon vertex\label{appendix:Feynman_rules}}

\begin{figure}[htbp]
  \begin{center}
    \begin{tabular}{c}
      \begin{minipage}{0.2\hsize}
        \begin{center}
          \includegraphics[clip, width=4.5cm]{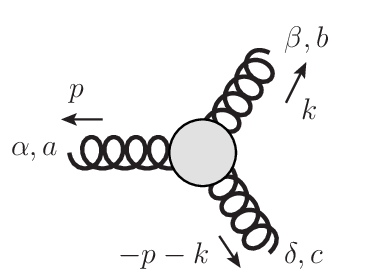}
          \vspace{3cm}
        \end{center}
      \end{minipage}
      \begin{minipage}{0.55\hsize}
        \begin{center}
\begin{eqnarray}
&=& 
i\frac{2}{3}w f_{abc}
i
\Biggl\{
\bigl[p^2 k_\gamma 
+k^2 p_\gamma 
+2(p\cdot k ) (p_\gamma +k_\gamma )
\bigr]
\nonumber\\
&& \hspace{8em} \times
\epsilon^{\alpha \beta \gamma \delta} 
\epsilon^*_\alpha (p)
\epsilon^*_\beta (k)
\epsilon^*_\delta (-p-k) 
\nonumber\\
&& \hspace{5em}-
\epsilon^{\alpha \beta \gamma \mu} 
\epsilon^*_\alpha (p)
\epsilon^*_\beta (k)
p_\gamma 
k_\mu
\,
(p^\delta-k^\delta) 
\epsilon^*_\delta (-p-k) 
\nonumber\\
&& \hspace{5em}-
\epsilon^{\delta \alpha \gamma \mu} 
\epsilon^*_\delta (-p-k)
\epsilon^*_\alpha (p)
p_\gamma 
k_\mu
\,
(-k^\beta-2p^\beta) 
\epsilon^*_\beta (k) 
\nonumber\\
&& \hspace{5em}-
\epsilon^{\beta \delta \gamma \mu} 
\epsilon^*_\beta (k)
\epsilon^*_\delta (-p-k)
p_\gamma 
k_\mu
\,
(p^\alpha+2k^\alpha) 
\epsilon^*_\alpha (p) 
\Biggr\}
\nonumber
\end{eqnarray}
          \hspace{1.6cm}
        \end{center}
      \end{minipage}

    \end{tabular}
    \caption{Feynman rule for the Weinberg operator.
    }
    \label{fig:wein_feyn}
  \end{center}
\end{figure}

We derive the Feynman rules of the three-gluon vertex generated by the Weinberg operator.
The three-gluon part of its Lagrangian is
\begin{eqnarray}
{\cal L}_w 
&=& 
\frac{1}{3!} w 
f^{abc} \epsilon^{\alpha \beta \gamma \delta} G^a_{\mu \alpha } G_{\beta \gamma}^b G_{\delta}^{\ \ \mu,c}
\nonumber\\
&=&
-\frac{1}{3} w
f^{abc} \epsilon^{\alpha \beta \gamma \delta} 
(\partial_\beta A_\gamma^b)
\Bigr[
   (\partial_\mu A_\alpha^a) (\partial^\mu A_\delta^c) 
-2 (\partial_\alpha A_\mu^a) (\partial^\mu A_\delta^c) 
 \Bigl]
+ O(A^4)
,
\label{eq:weinberg_three-gluon}
\end{eqnarray}
where we omitted the contribution from higher orders in the QCD coupling.
The Feynman rule for the Weinberg operator (\ref{eq:weinberg_three-gluon}) is shown in Fig. \ref{fig:wein_feyn}.
The greek letters $\alpha, \beta $, and $ \delta$ are the Lorentz indices of the gluon polarizations, and $a,b,c$ are the color ones. 

\section{Two-body amplitude (interquark potential)\label{appendix:2-body}}

The amplitude of the diagram of Fig. \ref{fig:2qint} is given by
\begin{eqnarray}
i{\cal M}
&=&
i\frac{2}{3}wf_{abc} \bar u_2 (p_2 +q) \gamma_\nu t_a u_2 (p_2) \frac{-g_s^3}{q^2 -m_g^2} 
\nonumber\\
&&  \times
\int \frac{d^4 k}{(2 \pi )^4}
\frac{1}{[k^2 -m_g^2] [ (k+p_1)^2-m^2] [(k+q)^2 -m_g^2]}
\nonumber\\
&& \hspace{2em} \times
\Biggl\{
\epsilon^{\nu \beta \gamma \delta} 
\bar u_1 (p_1 -q) \gamma_\delta t_c (k \hspace{-0.45em}/\,+p \hspace{-0.5em}/\,_1 + m ) \gamma_\beta t_b u_1 (p_1 )
 [q^2 k_\gamma +k^2 q_\gamma + 2(k\cdot q) (k_\gamma + q_\gamma )]
\nonumber\\
&& \hspace{4em} -
\epsilon^{\nu \beta \gamma \mu} q_\gamma k_\mu 
\bar u_1 (p_1 -q) (q \hspace{-0.5em}/\,-k \hspace{-0.45em}/\, ) t_c (k \hspace{-0.45em}/\,+p \hspace{-0.5em}/\,_1 + m ) \gamma_\beta t_b u_1 (p_1 )
\nonumber\\
&& \hspace{4em} -
\epsilon^{\delta \nu \gamma \mu} q_\gamma k_\mu 
\bar u_1 (p_1 -q) \gamma_\delta t_c (k \hspace{-0.45em}/\,+p \hspace{-0.5em}/\,_1 + m ) (-2 q \hspace{-0.5em}/\,-k \hspace{-0.45em}/\, )  t_b u_1 (p_1 )
\nonumber\\
&& \hspace{4em} -
\epsilon^{\beta \delta \gamma \mu} q_\gamma k_\mu (q^\nu +2 k^\nu )
\bar u_1 (p_1 -q) \gamma_\delta t_c (k \hspace{-0.45em}/\,+p \hspace{-0.5em}/\,_1 + m ) \gamma_\beta t_b u_1 (p_1 )
\Biggr\}
,
\end{eqnarray}
where the indices of the Dirac spinor $u$ label the quarks.
The loop integral is of course divergent due to the Weinberg operator which has a mass dimension of 6.
To remove the divergence, we apply the heavy quark approximation, which goes to the nonrelativistic quark model.
We then have
\begin{eqnarray}
i{\cal M}
&=&
\frac{N_c}{3}w \bar H_2 \gamma_\nu t_a H_2 \frac{-g_s^3}{q^2-m_g^2} 
\int \frac{d^4 k}{(2 \pi )^4}
\frac{1}{\bigl[k^2 - m_g^2\bigr] \bigl[(k+q)^2 - m_g^2\bigr] \bigl[ k\cdot v +i \epsilon \bigr] }
\nonumber\\
&& \hspace{10em} \times
\Biggl\{
\epsilon^{\nu \beta \gamma \delta} 
\bar H_1 \gamma_\delta {\cal P}_+ \gamma_\beta t_a H_1
 [q^2 k_\gamma +k^2 q_\gamma + 2(k\cdot q) (k_\gamma + q_\gamma )]
\nonumber\\
&& \hspace{12em} -
\epsilon^{\nu \beta \gamma \mu} q_\gamma k_\mu 
\bar H_1 (q \hspace{-0.5em}/\,-k \hspace{-0.45em}/\, ) {\cal P}_+ \gamma_\beta t_a H_1
\nonumber\\
&& \hspace{12em} -
\epsilon^{\delta \nu \gamma \mu} q_\gamma k_\mu 
\bar H_1 \gamma_\delta {\cal P}_+ (-2 q \hspace{-0.5em}/\,-k \hspace{-0.45em}/\, )  t_a H_1
\nonumber\\
&& \hspace{12em} -
\epsilon^{\beta \delta \gamma \mu} q_\gamma k_\mu (q^\nu +2 k^\nu )
\bar H_1 \gamma_\delta {\cal P}_+ \gamma_\beta t_a H_1
\Biggr\}
,
\label{eq:amplitude1}
\end{eqnarray}
where $v$ is the velocity 4-vector of the quark 1, defined by separating the scale of the 4-momentum of the quark as $p_\mu = m_Q v_\mu + k_\mu$, and $H_1 \equiv e^{im_Q x} {\cal P}_+ \psi (x)$.
Here the nonrelativistic projector of the quark ${\cal P}_\pm \equiv \frac{1}{2} (1\pm v\hspace{-0.45em}/\,)$ satisfies ${\cal P}_+ H_1=H_1$.

The point is to reduce the two-quark scattering amplitude of Eq. (\ref{eq:amplitude1}) to the CP-odd 2-body potential with the form $\vec{\sigma}\cdot \vec \nabla f(r) $, which is the leading contribution to the EDM in the nonrelativistic quantum mechanics.
Such a structure can be obtained from the color dipole moment-vector current interaction.
The loop integral of Eq. (\ref{eq:amplitude1}) must then give a chromo-EDM to contribute to the CP-odd potential.
Since the chromo-EDM involves a chirality flip, we take the scalar part of the projector ${\cal P}_\pm \equiv \frac{1}{2} (1\pm v\hspace{-0.45em}/\,) \to \frac{1}{2}$.
We also take the leading order terms in the exchanged momentum $q$, to agree with the nonrelativistic quark model.
The CP-odd two-quark scattering amplitude is then
\begin{eqnarray}
i{\cal M}
&=&
\frac{N_c}{6}w \bar H_2 \gamma_\nu t_a H_2 \frac{-g_s^3}{q^2 -m_g^2} 
\int \frac{d^4 k}{(2 \pi )^4}
\frac{1}{ \bigl[k^2 -m_g^2 \bigr] \bigl[(k+q)^2 -m_g^2 \bigr] \bigl[ k_0 +i \epsilon \bigr]}
\nonumber\\
&& \hspace{10em} \times
\Biggl\{
\epsilon^{\nu \beta \gamma \delta} 
\bar H_1 \gamma_\delta \gamma_\beta t_a H_1
 [q^2 k_\gamma +k^2 q_\gamma + 2(k\cdot q) (k_\gamma + q_\gamma )]
\nonumber\\
&& \hspace{12em} -
\epsilon^{\nu \beta \gamma \mu} q_\gamma k_\mu 
\bar H_1 (q \hspace{-0.5em}/\,-k \hspace{-0.45em}/\, ) \gamma_\beta t_a H_1
\nonumber\\
&& \hspace{12em} -
\epsilon^{\delta \nu \gamma \mu} q_\gamma k_\mu 
\bar H_1 \gamma_\delta (-2 q \hspace{-0.5em}/\,-k \hspace{-0.45em}/\, )  t_a H_1
\nonumber\\
&& \hspace{12em} -
\epsilon^{\beta \delta \gamma \mu} q_\gamma k_\mu (q^\nu +2 k^\nu )
\bar H_1 \gamma_\delta \gamma_\beta t_a H_1
\Biggr\}
\nonumber\\
&\approx &
\frac{N_c}{6}w \bar H_2 t_a H_2 \frac{-g_s^3}{q^2-m_g^2} 
\int \frac{d^4 k}{(2 \pi )^4}
\frac{({\rm Principal \ value})-i\pi \delta (k_0)}{\bigl[ k^2  -m_g^2 \bigr]^2 }
\nonumber\\
&& \hspace{10em} \times
\Biggl\{
\epsilon^{0 \beta \gamma \delta} 
\bar H_1 \gamma_\delta \gamma_\beta t_a H_1
 [k^2 q_\gamma + 2(k\cdot q) k_\gamma ]
\nonumber\\
&& \hspace{12em} 
+
\epsilon^{0 \beta \gamma \mu} q_\gamma k_\mu 
\bar H_1 k \hspace{-0.45em}/\, \gamma_\beta t_a H_1
+
\epsilon^{\delta 0 \gamma \mu} q_\gamma k_\mu 
\bar H_1 \gamma_\delta k \hspace{-0.45em}/\,  t_a H_1
\Biggr\}
\nonumber\\
&=&
i g_s^3\pi \frac{N_c}{6}w \bar H_2 t_a H_2 \frac{1}{q^2-m_g^2} 
\int \frac{d^4 k}{(2 \pi )^4}
\frac{\delta (k_0) k^2}{\bigl[ k^2 -m_g^2 \bigr]^2 }
\nonumber\\
&& \hspace{12em} \times
\Biggl\{
\frac{3}{2} q_\gamma 
\epsilon^{0 \beta \gamma \mu} 
\bar H_1 \gamma_\mu \gamma_\beta t_a H_1
+\frac{1}{2}
\epsilon^{0 \beta \gamma \mu} q_\gamma
\bar H_1 \gamma_\mu \gamma_\beta t_a H_1
\Biggr\}
\nonumber\\
&=&
-2i
g_s^3\pi \frac{N_c}{3}w \bar H_2 t_a H_2 \frac{1}{q^2} 
\bar H_1 
\sigma^{0 \gamma} q_\gamma \gamma_5
t_a H_1
\int \frac{d^4 k}{(2 \pi )^4}
\frac{\delta (k_0) k^2}{ \bigl[ k^2 -m_g^2 \bigr]^2 }
\nonumber\\
&=&
g_s^3 \frac{N_c}{3}w \bar H_2 t_a H_2 \frac{1}{q^2} 
\bar H_1 
i \sigma^{0 \gamma} q_\gamma \gamma_5
t_a H_1
\int \frac{d^3 \vec{k}}{(2 \pi )^3}
\frac{|\vec{k}|^2}{\Bigl[ |\vec{k}|^2 +m_g^2 \Bigr]^2 }
,
\label{eq:amplitude2}
\end{eqnarray}
where we used $\frac{1}{2} \epsilon^{0 \beta \gamma \mu} \sigma_{\gamma \mu} = -i \sigma^{0\beta} \gamma_5$.
The integral can be calculated with the dimensional regularization:
\begin{eqnarray}
&&
\int \frac{d^D \vec{k}}{(2 \pi )^D}
\frac{|\vec{k}|^2}{\Bigl[ |\vec{k}|^2 +m_g^2 \Bigr]^2 }
=
\frac{1}{(4 \pi)^{D/2}} \frac{D}{2} \frac{\Gamma (2-D/2-1)}{\Gamma (2) } \Biggl( \frac{1}{m_g^2} \Biggr)^{2-D/2-1}
\nonumber\\
&\to &
\frac{1}{(4 \pi)^{3/2}} \frac{3}{2} \Gamma (-1/2) \Biggl( \frac{1}{m_g^2} \Biggr)^{-1/2}
=
-\frac{3}{8 \pi }m_g
,
\end{eqnarray}
where we took the limit $D\to 3$.
Here we used $\Gamma(-1/2) = - \sqrt{4 \pi}$.
The final form of the two-body scattering amplitude is
\begin{eqnarray}
i{\cal M}
&=&
-
\frac{N_c g_s \alpha_s m_g }{2 }
w
\frac{1}{q^2 - m_g^2} 
\bar H_2 t_a H_2 
\cdot
\bar H_1 
i \sigma^{0 \gamma} q_\gamma \gamma_5
t_a H_1
\nonumber\\
&\approx &
\frac{N_c g_s \alpha_s m_g}{2 }
w
\frac{1}{|\vec{q}|^2 + m_g^2} 
\bar H_2 t_a H_2 
\cdot
\bar H_1 
t_a \vec{\sigma}\cdot \vec{q}
H_1
.
\end{eqnarray}
By adding the contribution from the diagram with the quarks interchanged, we obtain Eq. (\ref{eq:amplitude3}).

\end{document}